\documentclass[3p,times,procedia]{elsarticle}
\usepackage{ecrc}
\usepackage[bookmarks=false]{hyperref}
    \hypersetup{colorlinks,
      linkcolor=blue,
      citecolor=blue,
      urlcolor=blue}

\volume{00}

\firstpage{1}

\journalname{Procedia Computer Science}

\runauth{Tellez-Ballesteros et al.}

\jid{procs}

\usepackage{amssymb}

\usepackage[figuresright]{rotating}

\begin{document}
\begin{frontmatter}



\dochead{6th International Conference on Industry 4.0 and Smart Manufacturing}

\title{Sales predictive analysis for improving supply chain drug sample}


\author[a]{Susana Casy Téllez-Ballesteros\corref{cor1}} 
\author[a]{Ricardo Torres-Mendoza}
\author[a]{José Antonio Marmolejo-Saucedo}
\author[b]{Roman Rodriguez-Aguilar}

\address[a]{Facultad de Ingenieria, Universidad Nacional Autonoma de Mexico, Av. Universidad 3000, 04510, Mexico City, Mexico}
\address[b]{Facultad de Ciencias Económicas y Empresariales, Universidad Panamericana, Augusto Rodin 498, 03920, Mexico City, Mexico}

\begin{abstract}
The delivery of drug samples allows increasing sales of pharmaceutical products \cite{Dong2013}. However, we discovered some problems that can be improved in the supply chain that delivers drug samples (used for the treatment of excess glucose). Databases were integrated; then we apply data extraction and transformation; and finally we apply multiple regression analysis to explain drug sales. The first analysis evaluates the integration of regional data and the second analysis refers to data dis-aggregated by region. We identify the region with the greatest impact on sales and the impact of the delivery of drug samples in the Mexican market.
\end{abstract}

\cortext[cor1]{Corresponding author. Tel.: +015-55622-9983.}
\end{frontmatter}

\email{stellezb@unam.mx}



\section{Introduction}
\label{main}

A drug sample is the presentation of a pharmaceutical product that will be provided free of charge in accordance with the Law and regulations issued by the Health Authority of a country. A drug sample of a pharmaceutical product is provided to doctors to advertise the product. Samples must be labeled as such, so they cannot be resold or have an inappropriately used. There are other disadvantages in the delivery of medical samples, and original gifts, because it is a reduced presentation of the product pharmaceutical, which involves enormous amounts of money.
\label{main}

In the distribution of drug samples in hospitals and clinics, the Law must be respected General Health, its regulations, agreements and standards established for this purpose. The delivery of drug samples is a marketing and promotion strategy that pharmaceutical companies carry out as an alternative to promote their product. This type of promotion is carried out with the purpose of giving get to know the products to professionals in the area, increase the numbers of their sales in a new product and quickly position it in the market. However, there are also disadvantages in drug sample delivery, because it is a reduced presentation of the product pharmaceutical, which involves huge amounts of money.

First paper part presents a review of the analysis of sample delivery in other countries. Second paper part is the description of the project in a Mexican pharmaceutical company between September 2018 and March 2020. The objective of the project was to know delivery drug samples impact on pharmaceutical drug sales. In the last paper part we describe our conclusions through data analysis and multiple regression methodologies.

\section{Literature review}
\label{main}
Drug samples as promotional tools outperform other promotional inputs offered to doctors, and it is a costly promotional input for companies \cite{Bala2013} and \cite{Harindranath2017}. The objective of medical samples is to give presence to the product and build loyalty of the Doctor's prescription to his patients \cite{Chew2000}  ,\cite{Gönül2012}, \cite{Lim2008}, and \cite{Mizik2004}. When a doctor becomes familiar with a medication, he is more likely to prescribe it compared to others \cite{Schramm2009}. Also, medication samples allow patients to start treatment and evaluate its effectiveness \cite{Groves2003}. If the medication turns out to improve the patient's conditions, it will be a predictor of future consumption \cite{Schramm2009}.

\label{main}

In addition to the initiation of trials, sampling serves other purposes, such as the introduction of new and/or unique products, the displacement of an entrenched market leader \cite{Groves2003}. Even more, if the medical sample allows the evaluation of an effective drug compared to an ineffective one, as well as the presence of side effects in its consumption, it will facilitate the increase in demand for the drug in the market \cite{Venkataraman2007}.

\label{main}

The delivery of medical samples makes it possible to pay for the treatment of patients with lower purchasing power \cite{Chew2000}. Providing free samples to price-sensitive patients can improve relationships between doctors and patients \cite{Gönül2012}. Therefore, drug samples can act in two opposite ways: they can increase sales by stimulating trials or they can cannibalize sales \cite{Dong2013}, suggesting that the Pharmaceutical Sales Representatives (PSR) should be careful during drug distribution. the drugs. samples, change the image of a product or even quickly build a franchise or consumer demand \cite{Groves2003}.

\label{main}

The academic affiliation of the doctor negatively influences the interaction with representatives of the medical sector. Restrictive physician contact policies with medical representatives. Primary care physicians may be more receptive to obtaining medication information from sources that include PSR. Physicians with a lower prescribing profile interact with medical representatives \cite{Alkhateeb2009} and 
\cite{Bennett2005}. Physicians who work in small practices are more likely to see PSR, compared to those who work in larger practices \cite{Alkhateeb2009} and \cite{Andaleeb1995}.

\label{main}

Other studies of drug samples such as their unregulated manipulation, self-medication by health professionals and representatives of the pharmaceutical industry, unregulated use and the elimination of unused and/or expired samples.\cite{Groves2003} and \cite{Schramm2009}.

\label{main}

Predictive analytics is an operational analytical tool of artificial intelligence described artificial intelligence as a technology that builds Pharma 4.0 as an Industry 4.0 healthcare application \cite{Harindranath2017}. However, they only conceptualize Pharma 4.0 technologies, they do not describe their applications. Table 1 shows paper with perspective analysis of the impact of drug sample.

\label{main}

Nguyen concluded that COVID-19 pandemic created an urgent need to investigate data analysis and artificial intelligence capabilities to provide support in times of crisis \cite{Nguyen2021}. Particular at the sample drugs supply distribution strategy, the follow section is about a Mexican pharmaceutical, and their interest to knowing about the influence of samples drugs.  

\begin{table}[h]
\caption{Research paper with perspective analysis}
\begin{tabular*}{\hsize}{@{\extracolsep{\fill}}lll@{}}
\toprule
Author, year & Country & Methodology \\
\colrule
Venkataraman et. Al., 2007 &   United States &  Data bases and econometric model\\
Alkhateeb, et. Al., 2009  & United States  &  Surveys, and regression\\
Schramm, et. Al. 2009 &  Denmark &  Statistics record of marketing\\
Harindranath, et. Al. 2017 &  India &  Surveys, and regression\\
Soharabi, et. Al, 2019 &  Iran &  Surveys, and correlation analysis\\
Tellez-Ballesteros, et. Al, 2024 &  Mexico &  Analytics and correlation analysis\\
\botrule
\end{tabular*}
\end{table}

\section{Case Study}
\label{main}

The drug sample that we are going to evaluate is a medication dedicated to the treatment of a chronic metabolic disease in which excess glucose or sugar is produced in the blood. The study is carried out based on historical data (September 2018 to March 2020) on sales, medical prescriptions, and delivered samples of the medication being studied. The objective of this research is to know if there is an impact on sales due to the delivery of drug samples. Due to confidentiality agreements, we cannot reveal the source of the data. The data set allows knowing delivery region, number of drug samples and delivery date. Finally, we build data on the target population of this treatment by region and date.

\label{main}

The process began with the integration of the databases on the variable of interest. We extracted data sets of relevant information, then transformed the data set with the same periodicity, and then loaded the final data. We tested several models: the first analysis evaluates the integration of regional data. and the second analysis dis-aggregated data evaluated by region.  Finally, we report the results and conclusion, see Fig.~1.

\begin{figure}[t]\vspace*{4pt}
\centerline{\includegraphics[height=3cm]{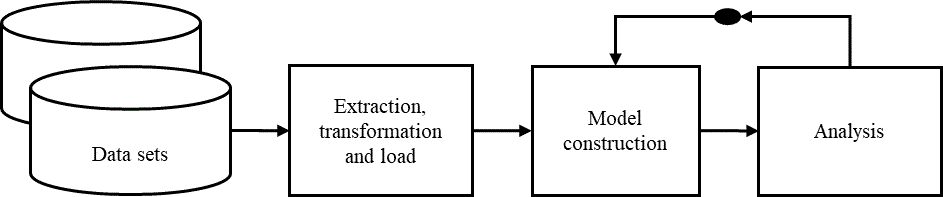}}
\caption{Process project, Source: own elaboration.}
\end{figure}

\label{main}

To estimate the effects of delivery sample drugs at the sales treatment drugs, we use the follow regression analysis (see equation~1)

\begin{equation}
\begin{array}{lcl}
\displaystyle Y &=& \displaystyle\beta_{0}+\beta_{1}{X}_{1} +\beta_{2}{X}_{2} +\beta_{3}{X}_{3} \\[6pt]
\end{array}
\end{equation}

\begin{nomenclature}
\begin{deflist}[A]
\defitem{$Y$}\defterm{is dependent variable (sales treatment drug)}
\defitem{$X_{1}$}\defterm{is independent variable (period)}
\defitem{$X_{2}$}\defterm{is independent variable (target population)}
\defitem{$X_{3}$}\defterm{is independent variable (number or sample drugs deliver) }
\defitem{$\beta_{0}$}\defterm{is the intercept term }
\defitem{$\beta_{i} $}\defterm{for i= 1, 2, 3; are regression coefficients for the independent variables }
\end{deflist}
\end{nomenclature}

\subsection{ Analysis with integration of regional data}
Multiple regression analysis describes the influence of a set of independent variables on a dependent variable. It allows you to evaluate what happens to the dependent variable when an independent variable is changed. We tasted two model with data integration of independent and depend for all regions.We follow the follow methodology:
\begin{itemize}[]
\item First: It is about selecting the model to evaluate
To estimate the effects of delivery sample drugs at the sales treatment drugs, we use the follow regression analysis (see equation~2).
\begin{equation}
\begin{array}{lcl}
\displaystyle Sales &=& \displaystyle\ {23352.23726} + {174.1386462} * {Period} - {0.049844533} * Population	\\[6pt]
\end{array}
\end{equation}
\begin{figure}[t]\vspace*{4pt}
\centerline{\includegraphics[height=5cm]{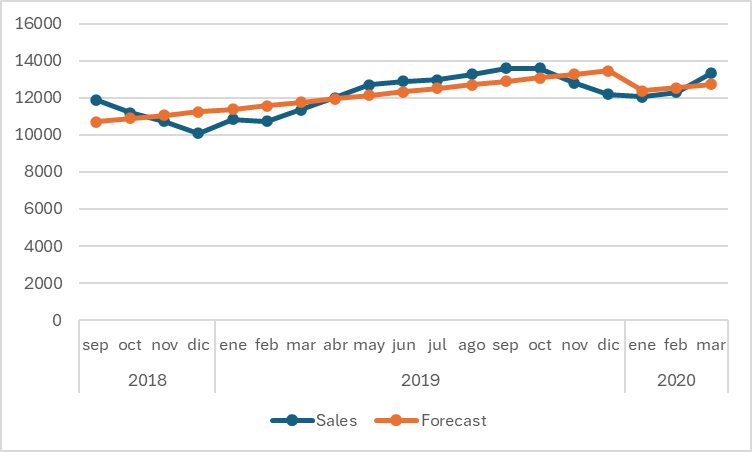}}
\caption{Forecast with regression model for all region vs sales treatment drug . Source: own elaboration.}
\end{figure}

\item Second: It consists of performing the F-test, which has the following null hypothesis: “the regression model does not explain the dependent variable” with a significance level of {$\alpha = {0.10.}$}
The Analysis of variance (ANOVA) let us select better model with a significance of regression with F-test value of $3.16*10^{-12}$. We reject the null hypothesis that regression model does not explain the depend variable because F-test is less than 0.10.
\
\item Third: It consists of performing the t-test. The null hypothesis is to test that each independent variable does not contribute to the explanation of the dependent variable with a significance level of {$\alpha = {0.10}$}. We see at table~2 that probability is less than $0.1$; therefore, we reject the null hypothesis that each partial regression coefficient is equal to zero and conclude that each of them is statistically significant.

\begin{table}[h] \caption{Variables t-test.}
\begin{tabular*}{\hsize}{@{\extracolsep{\fill}}lll@{}}
\toprule
Variable & t statistician & Probability\\
\colrule 
Interception &  3.93940 & 0.00117 \\
Period & 4.76771 &	0.00020 \\
Population  & -2.11802 & 0.05017 \\
\botrule
\end{tabular*}
\end{table}

\item Fourth. Then the coefficient of determination $R^{2}$ is evaluated. This indicates how much of the change in the Y variable is explained by the regression model. We are obtaining a coefficient of determination equal to $0.59$. Since its value is greater than 0.3, it might be considered acceptable.
\end{itemize}
We estimate the regression function using the ordinary least squares method, this involves evaluating the following.
\begin{itemize}[]
\item First assumption. We are drawing the scatter plots for all metric independent variables and then visually inspect whether there is a linear relationship between the respective independent variable and the dependent variables, see figure~3.
\begin{figure}\vspace*{4pt}
\centerline{\includegraphics[height=4cm]{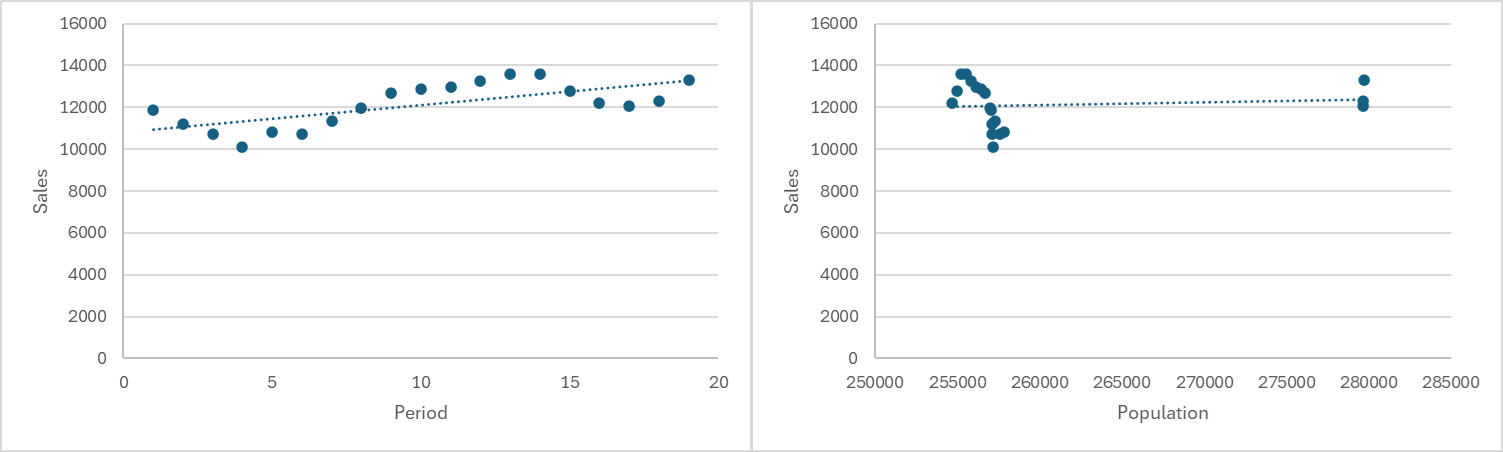}}
\caption{Sales vs independent variables}
\end{figure}
\item Second assumption. It requires that the deviation of the observed values (Sales) from the estimated values (Forecast sales) have an expected value of zero. We are obtaining deviation equal to 681.
\item Third assumption. It demands that the deviations are not correlated with the independent variables. We can evaluate this again with the scatter plot. We plot the residuals or the standardized residuals against the independent variables at figure~4. We can not recognized trend or pattern at the graph.
\begin{figure}\vspace*{4pt}
\centerline{\includegraphics[height=4cm]{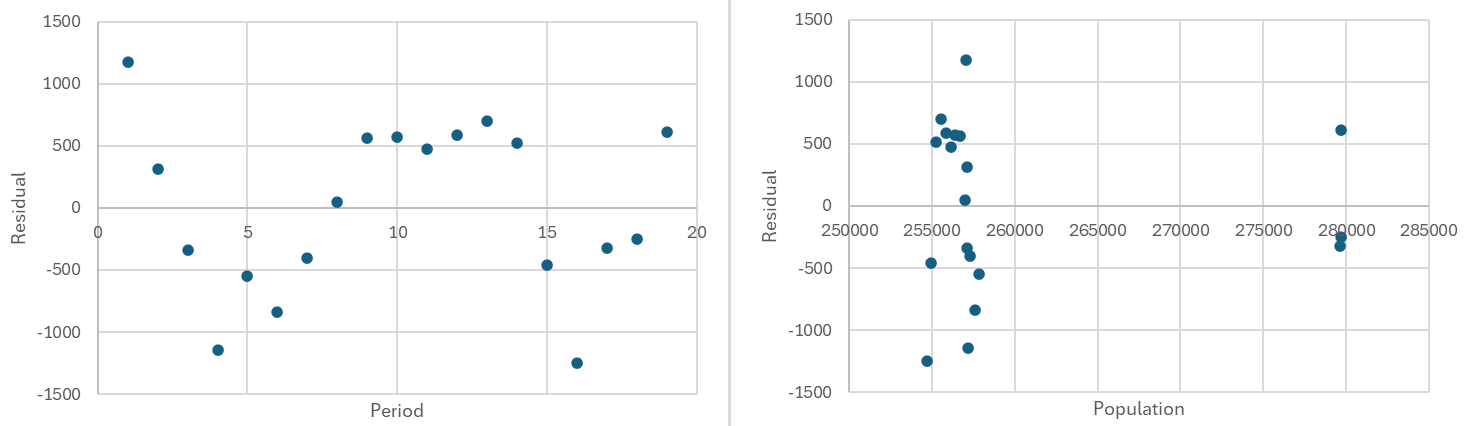}}
\caption{Residual vs independent variables}
\end{figure}
\item Fourth assumption.  It is about evaluating homoscedasticity. This means that the variance is constant over the range of the estimated dependent variable (sales forecast), Since we cannot see the trend or pattern in Figure~5, it means that the variance is constant.
\begin{figure}\vspace*{4pt}
\centerline{\includegraphics[height=4cm]{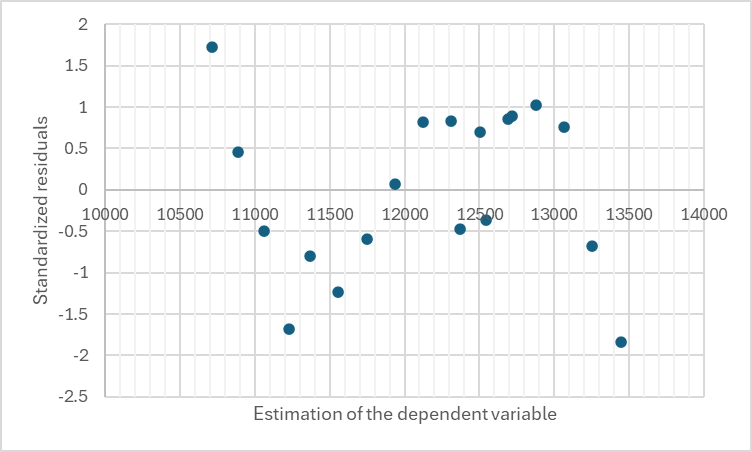}}
\caption{Standardized residuals vs estimation of the dependent variable}
\end{figure}
\item Fifth assumption. It is evaluating that independent variables are not correlated with each other. As we can see in table ~3, each correlation is less than 0.8, this means that there are no correlated variables.
\begin{table}[h] \caption{Correlation of independent variables}
\begin{tabular*}{\hsize}{@{\extracolsep{\fill}}llll@{}}
\toprule
Variable & Period & Population & Sales \\\colrule 
Period & 1&		\\
Population & 0.55 &	1	\\
Sales & 0.69 & 0.10 & 1 \\
\botrule
\end{tabular*}
\end{table}
\item Sixth assumption. This assumption requires that the deviations are not correlated with each other. We must plot the observations in sequence on the x-axis and the residuals on the y-axis. If a pattern is not recognized in the residuals, the assumption is not violated (see figure~6a).
\begin{figure}\vspace*{4pt}
\centerline{\includegraphics[height=4cm]{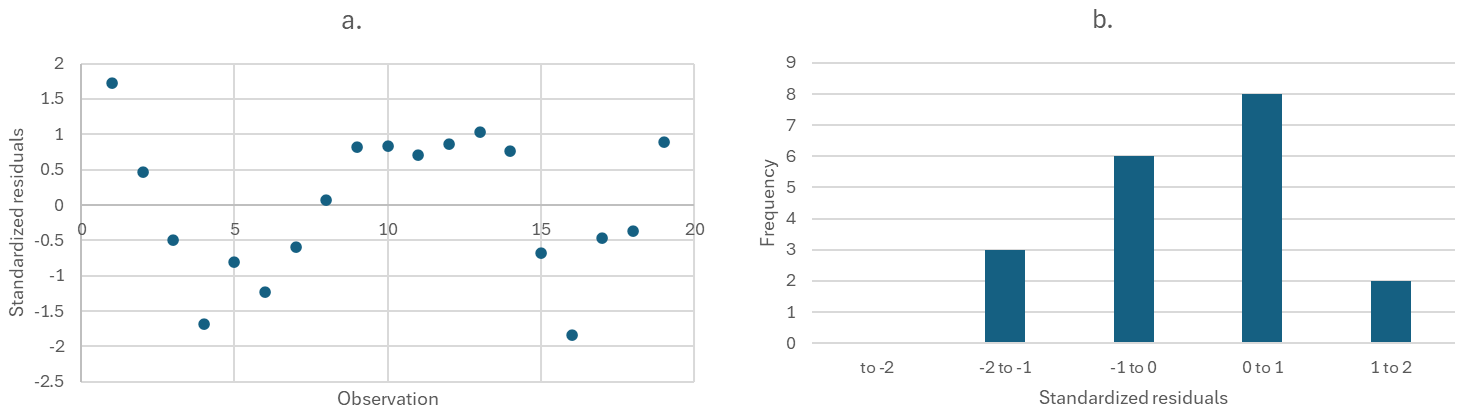}}
\caption{(a) Standardized residuals vs observation; (b) Examination of the normal distribution of residual}
\end{figure}
\item Seventh assumption. It demands that the residuals are normally distributed. We are evaluating this requirement visually with the histogram in figure ~6b.
\end{itemize}

\label{main}
\subsection{Analysis with dis-aggregated data by region }
The target of this analysis is to develop a strategy for delivering sample drugs by region. We tested several models with region data and finally selected the best model for each. Four columns are shown in table~5, the first refers to the region, the second column shows the better function by region, the third column shows the calculation of the coefficient of determination and the last column shows the F-test calculation.
\begin{table}[h] \caption{Function Sales by region}
\begin{tabular*}{\hsize}{@{\extracolsep{\fill}}llll@{}}
\toprule
State & Function & Coefficient of determination & F-test\\
\colrule 
CAM & Sales CAM = 10.64 - 0.016 * Sample	& \textbf{0.09}	& \textbf{0.70579} \\
MEX	& Sales MEX = 766.391 + 0.09 * Sample	& 0.31	& \textbf{0.1999}\\
MIC	& Sales MIC = 271.843 - 0.151 Sample	& \textbf{0.25}		& \textbf{0.29478} \\
SON	& Sales SON = 60.951 + 0.077 Sample	& 0.33	& \textbf{0.16524}\\
ZAC	& Sales ZAC = 31.14 + 0.233 Sample	& 0.64	& 0.00327\\
AGU	& Sales AGU = 1325.308 - 0.371 Population	& 0.47	& 0.04118 \\
CHH	& Sales CHH = 1204.753 - 0.393 Population	& 0.48	& 0.03553\\
ROO	& Sales ROO = -31.372 + 0.014 Population	& 0.57	& 0.01052\\
TAB	& Sales TAB = 296.249 - 0.036 Population	& \textbf{0.22}	& \textbf{0.35796}\\
TAM	& Sales TAM = 739.279 - 0.01 Population	& \textbf{0.29}	& \textbf{0.23386}\\
BCN	& Sales BCN = 268.588 + 23.745 Period	& 0.6	& 0.02825\\
CHP	& Sales CHP = 80.368 + 3.247 Period	& 0.8	& 0.00005\\
CMX	& Sales CMX = 2936.947 + 41.979 Period	& 0.62	& 0.00474\\
COA	& Sales COA = 274.684 + 1.368 Period	& \textbf{0.17} & \textbf{0.49112}\\
DUR	& Sales DUR = -852.216 - 4.39 Period & 0.74	& 0.00175\\
GUA	& Sales GUA = 1362.212 + 4.895 Period & 0.64 & 0.01426\\
GRO	& Sales GRO = 73.795 + 7.591 Period	& 0.77 & 0.00083\\
HID	& Sales HID = 61.877 + 0.244 Period	& \textbf{0.14} & \textbf{0.57492}\\
JAL	& Sales JAL = 1765.877 + 24.281 Period & 0.86 & 0.00003\\
MOR	& Sales MOR = 70.07 + 2.161 Period	& 0.68	& 0.00147\\
NAY	& Sales NAY = 126.158 - 2.726 Period & 0.72	& 0.0005\\
NLE	& Sales NLE = 629.965 + 2.893 Period & \textbf{0.23}	& \textbf{0.35431}\\
OAX	& Sales OAX = 60.246 + 5.891 Period	& 0.82	& 	0.00002\\
PUE	& Sales PUE = 229.4 - 4.979 Period	&  0.53	& \textbf{0.16166}\\
QUE	& Sales QUE = 83.737 + 2.563 Period	&  0.78	& 0.00009\\
SIN	& Sales SIN = -1142.416 + 20.336 Period	& 0.94	& 3.82 $*10^{-8}$\\
SLP	& Sales SLP = 101.316 - 1.563 Period	& 0.56	& 	0.01184\\
VER	& Sales VER = 393.263 + 8.616 Period	& 0.65 & 0.00274\\
YUC	& Sales YUC = 90.86 + 6.919 Period	& 0.71	& 	0.00063\\
\botrule
\end{tabular*}
\end{table}

\label{main}
Table 5 shows the tested results of the initial model by region and by selecting the better fit, we obtain F-test value less than 0.10 for 19 states in the country of Mexico. In the same table it can be concluded that the first five states show that sales depend on drug samples, for two of them the coefficient is negative, that is, if the pharmacy delivers more drug samples then sales decrease (Campeche and Michoacan states). 
For 19 states, sales depend on the time period. For four states the coefficient for the period is negative, meaning that sales decrease if time increases (Durango, Nayarit, Puebla and San Luis Potosí states).

\label{main}
We integrated the forecast for each region and compared it with the actual value (see figure~7). 
\begin{figure}\vspace*{4pt}
\centerline{\includegraphics[height=4cm]{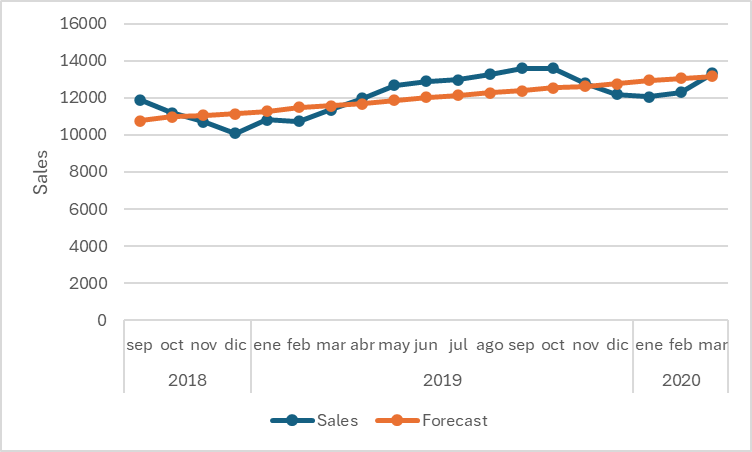}}
\caption{Forecast with regression model for disa-aggegated data by region vs sales treatment drug}
\end{figure}
\label{main}

\label{main}
We calculated the mean absolute percentage deviation (MAPE) of the sales forecast per region equal to 5.56\%.\ Compared to the absolute deviation of the sales forecast from the previous analysis equal to 4.9\%.\ We conclude that the delivery of drug samples is showing a minimal impact on sales, in both analyses.

\label{main}The pareto analysis for the sales region is shows at Figure~8. We related the sales states that represent the 79\%\ of the total sales in study period time with their forecast regression model.Of the ten states that represent 79\%\ of sales, only Estado de Mexico is explained by the delivery of samples.
\begin{figure}\vspace*{4pt}
\centerline{\includegraphics[height=4cm]{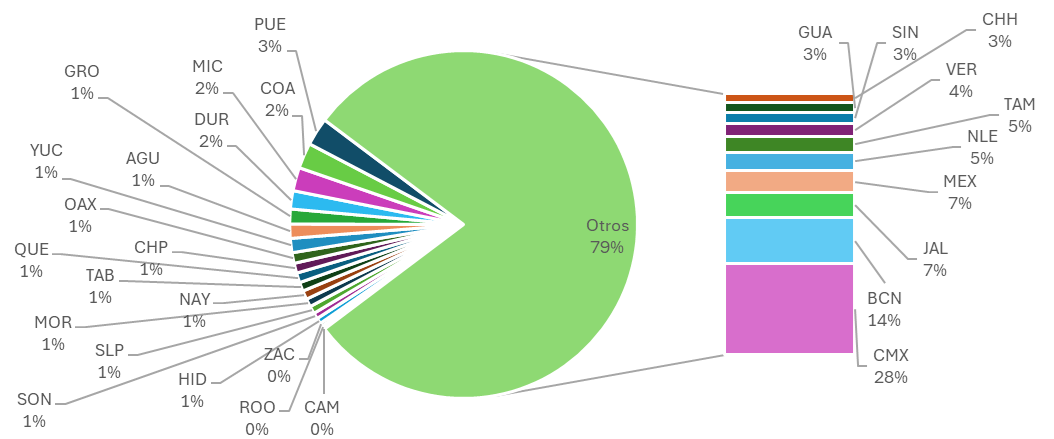}}
\caption{Pareto Analysis Sales by region}
\end{figure}
\label{main}
\section{Conclusions}
From both analyses, the sale of medications is not explained by the delivery of a sample of medication. In the model that explains the sale of medicines, the majority of sales by state are explained by time period. 

\label{main}
For the models by region that are explained by the delivery of samples and its coefficient is negative. In Campeche the delivery of samples exceeds sales, which suggests cannibalization of the market. 

\label{main}
We consider it advisable to carry out a subsequent study that evaluates the impact of the delivery of samples in periods after delivery. As well as the percentages of samples that are delivered with respect to the total sale, as another indicator of the commercial strategy.






\normalMode
\end{document}